\documentclass[twocolumn]{aastex631}

\usepackage{CJK}
\usepackage{multirow}
\usepackage[T1]{fontenc}

\begin{document}
\begin{CJK*}{UTF8}{gbsn}
\title{Comparing observed properties of winds in low-luminosity active galactic nuclei with theoretical predictions}

\author[0000-0003-3922-5007]{Fangzheng Shi (施方正)}
\affiliation{Shanghai Astronomical Observatory, Chinese Academy of Sciences, Shanghai, People's Republic of China}\email{fzshi@shao.ac.cn}

\author[0000-0003-3564-6437]{Feng Yuan}
\affiliation{Center for Astronomy and Astrophysics and Department of Physics, Fudan University, Shanghai 200438,
People's Republic of China}\email{fyuan@fudan.edu.cn}

\author{Francesco Tombesi}
\affiliation{Physics Department, Tor Vergata University of Rome, Via della Ricerca Scientifica 1, 00133 Rome, Italy}
\affiliation{INAF – Astronomical Observatory of Rome, Via Frascati 33, 00040 Monte Porzio Catone, Italy}
\affiliation{INFN - Rome Tor Vergata, Via della Ricerca Scientifica 1, 00133 Rome, Italy}

\author{Fu-guo Xie}
\affiliation{Shanghai Astronomical Observatory, Chinese Academy of Sciences, Shanghai, People's Republic of China}

%\collaboration{20}{(AAS Journals Data Editors)}

%\author{F.X Timmes}
%\affiliation{Arizona State University}
%\affiliation{AAS Journals Associate Editor-in-Chief}

%\altaffiliation{AASTeX v6+ programmer}



\begin{abstract}
Theoretical and numerical simulations of black hole hot accretion flows have shown the ubiquitous existence of winds and predicted their properties such as velocity and mass flux. In this paper, we have summarized from literature the physical properties of winds launched from low-luminosity active galactic nuclei (LLAGN), which are believed to be powered by hot accretion flows, and compared them with theoretical predictions.
We infer that for both ultra-fast outflows and hot winds, 
the observed wind velocity as a function of their launching radius and the ratio between wind mass flux and black hole accretion rate show good consistency with theoretical predictions.
For the prototype LLAGN M81* with abundant observational data, we have examined various observed properties of wind in detail, including velocity, mass flux of the wind, the power-law index of the radial profile of inflow rate, and the jet-to-wind power ratio.
Good agreements are found with theoretical predictions, providing strong support to the theory of wind launched from hot accretion flows.
\end{abstract}



\section{Introduction} \label{sec:intro}
Black hole accretion can be divided into cold mode and hot mode according to the mass accretion rate.
The cold mode accretion includes standard thin disks and super-Eddington accretion \citep{1973A&A....24..337S,1988ApJ...332..646A}.
It usually happens in black hole with high or intermediate accretion rate, like in bright quasars and Seyfert galaxies. 
While we focus on hot mode accretion where Eddington ratio (between AGN bolometric luminosity and the Eddington limit) falls below the threshold of 2\% \citep{2014ARA&A..52..529Y}.
Such black holes with low accretion rate would manifest themselves as low-luminosity AGN (LLAGN).
Majority of the local galaxy are in low-luminosity and quiescent states \citep{2008ARA&A..46..475H}.
As an analog of black hole binary (BHB) in low-hard state, LLAGN is found to have similar accretion structure. As the accretion rate of black hole drops, the inner accretion thin disk would be truncated and replaced by geometrically thick, optically thin, radiatively inefficient hot accretion flow \citep{2014ARA&A..52..529Y,1994ApJ...428L..13N,1995ApJ...452..710N,1997ApJ...489..865E}.

Both theory study and MHD numerical simulations of black hole accretion in the past decade have predicted the existence of strong wind launched from hot accretion flows \citep{1999MNRAS.303L...1B,2012ApJ...761..129Y,2012ApJ...761..130Y,2012MNRAS.426.3241N,2015ApJ...804..101Y,2016ApJ...818...83B}. 
The detailed properties of the hot wind, such as the velocity, mass flux, and special distribution have been investigated based on three-dimensional GRMHD simulations\citep{2015ApJ...804..101Y,2021ApJ...914..131Y}.
Hot wind not only affects the dynamics of accretion inflow around black hole, but also serves as an important medium in AGN feedback \citep{2012ARA&A..50..455F,2013ARA&A..51..511K,2017MNRAS.465.3291W,2018ApJ...857..121Y,2019ApJ...885...16Y}.
By including the hot wind feedback in the numerical simulation study of the evolution of an individual elliptical galaxy, it has been shown that wind plays an important role in controlling the luminosity of black hole and its growth \citep{2019ApJ...885...16Y}.

Observational detection of wind driven by hot accretion flow, however, remains scarce and indirect for a long time.
\citet{2013Sci...341..981W} reported a flat rather than increasing density profile of hot accretion flow around Sgr A* from high-resolution X-ray spectroscopy, implying indirectly the existence of potential outflow suppress the inflow.
\citet{2016Natur.533..504C} inferred the presence of centrally-driven, galactic-scale wind in a sample of typical quiescent galaxies hosting low-luminosity AGN around $z\sim0.02$.
The first direct observational evidence of hot wind driven by hot accretion flow in LLAGN has been detected through blue-shifted highly ionized iron emission lines in M81* and NGC\,7213. The kinetic energy carried by these LLAGN winds can account for 10\% -- 15\% of total AGN bolometric luminosity \citep{2021NatAs...5..928S,2022ApJ...926..209S}.
\textbf{This is larger than the theoretical prediction that the mechanical efficiency generated by the hot accretion flows can hardly exceed 3\% of the accretion mass energy flux claimed by \citet{2017MNRAS.468.1398S}.}
The observational evidence for the interaction between wind and the cicumnuclear gas within pc-scale has been found and is likely the physical mechanism of keeping  the central AGN dim \citep{2024ApJ...970...48S}.

Ionized winds characterized by blue-shifted absorption lines in X-ray or UV bands are ubiquitous in up to $30$\% -- $50$\% of local (bright) AGN\citep{2012ApJ...753...75C}.
Some ultra-fast outflows (UFO) could reach $\sim10^4\rm~km~s^{-1}$ while most warm absorbers (WA) have a velocity ranging from $\sim10^2$ to $\sim10^3\rm~km~s^{-1}$.
Many statistical studies have been carried on the whole sample or certain types of outflows in AGN \citep{2014MNRAS.441.2613L,2016MNRAS.457.3896L,2017A&A...601A.143F,2019A&A...625A..25M}. 
By comparing UFOs in both radio-loud and radio-quiet AGN, \citet{2024MNRAS.532.3036M} suggests that these accretion disk winds could likely be produced by the same physical mechanism.
More and more evidence support the driving mechanism (specifically for UFOs) would be dominated by magnetically driven in the inner region of accretion flow \citep[e.g.,][]{2015ApJ...804..101Y,2021ApJ...914..131Y,2022MNRAS.513.5818W,2022ApJ...940....6F}, and wind properties would depend on the accretion rate, black hole spin and magnetic field configuration and strength \citep{2015ApJ...804..101Y,2021ApJ...914..131Y}.
The Eddington ratio of AGN hosting either UFO or WA spans a very wide range. Some of these AGN has an Eddington ratio below the threshold of 2\%, where hot mode accretion should dominate their inner accretion region according to the theory of black hole accretion \citep{2014ARA&A..52..529Y}.
But systematic study of such winds detected in low-luminosity AGN remains scarce so far.

The aim of this work is to statistically summarize the common properties of outflows detected in LLAGN so far and to compare the results with the theory of wind launching in hot accretion flows.
In Sec.\ref{sec:method}, we first review the theoretical prediction of wind launched from hot accretion flows, then select a sample of outflows detected in the low-luminosity AGN with Eddington ratio lower than the critical value $R_{\rm Edd}\lesssim0.02$.
Statistical results of winds in these LLAGN have been reported in Sec.\ref{sec:stat}.
In Sec.\ref{sec:M81case}, we analyze the observed wind velocity, mass flux of wind, and inflow of M81* in detail and compare them with the state-of-the-art theory of hot accretion flow wind.
We summarize and discuss our results in Sec.\ref{sec:sum}.

\section{Method} \label{sec:method}
\subsection{Physics of hot accretion flow}

In LLAGNs, the accretion flow consists of an outer truncated thin disk plus an inner hot accretion flow \citep{1997ApJ...489...87S,2004ApJ...612..724Y,2014MNRAS.438.2804N,2018MNRAS.476.5698Y}.
The truncation radius ($r_{\rm tr}$) is inversely correlated with the Eddington ratio (or the accretion rate) \citep{2014ARA&A..52..529Y}. In the present paper we focus on the wind launched from the inner hot accretion flow. While the existence of such a wind has been shown by some previous works \citep[e.g.,][]{2012ApJ...761..130Y,2012MNRAS.426.3241N}, they did not investigate its physical properties. Since the accretion flow is highly turbulent, to understand the wind properties, it is crucial to separate the real wind from the turbulent outflow. In literature, a widely adopted approach is to calculate a time average of the the simulation data to filter out turbulence. However, since the wind is instantaneously produced, in this case, the real wind may also be filtered out. 

To overcome this difficulty, \citet{2015ApJ...804..101Y} adopted a new ``virtual test particle trajectory'' approach. Different from the widely adopted time-averaged streamline approach, obtaining Lagrangian trajectories
can adequately reflect the motion of fluid elements thus discriminate between wind and turbulence. Using this approach, based on three-dimensional general relativistic MHD numerical simulation data of black hole accretion, \citet{2015ApJ...804..101Y} successfully obtained the properties of the wind launched from a hot accretion flow. This work only deals with the SANE (standard and normal evolution) around a non-spinning black hole (SANE00). Later, \citet{2021ApJ...914..131Y} has extended this work to cases of both SANE and MAD (magnetically arrested disk)  around black holes with various spin values (i.e., SANE98, MAD00, and MAD98). Here the number denotes the spin value of the black hole. For example, ``98'' denotes $a=0.98$.

The net mass accretion rate in hot accretion flow would decrease with the decreasing of radius as a result of the presence of the hot wind.
The radial profile of the mass flux of the inflow  $\dot{M}_{\rm in}$ can be well described as:
\begin{equation}\label{eq:RadMin}
    \dot{M}_{\rm in}(r)=\dot{M}_{\rm in}(r_{\rm out})\left(\frac{r}{r_{\rm out}}\right)^s
\end{equation}
Here $r_{\rm out}$ denotes the outer boundary of the hot accretion flow. The values of  $s$-index are \citep{2015ApJ...804..101Y,2021ApJ...914..131Y}:
\begin{itemize}
    \item{s=0.54 for SANE00}
    \item{s=0.91 for SANE98}
    \item{s=0.18 for MAD00}
    \item{s=0.42 for MAD98}
\end{itemize}

For a wind launched from a hot accretion flow, the wind velocities are predicted to be:
\begin{equation}\label{eq:vSANE00}
    v_{\rm wind}=0.21v_{\rm k}(r_{\rm tr})~(\rm SANE00)
\end{equation}
\begin{equation}\label{eq:vSANE98}
    v_{\rm wind}=0.66v_{\rm k}(r_{\rm tr})~(\rm SANE98)
\end{equation}
\begin{equation}\label{eq:vMAD00}
    v_{\rm wind}=0.24v_{\rm k}(r_{\rm tr})~(\rm MAD00)
\end{equation}
\begin{equation}\label{eq:vMAD98}
    v_{\rm wind}=0.64v_{\rm k}(r_{\rm tr})~(\rm MAD98)
\end{equation}
Here $v_{\rm k}(r)$ refers to the Keplerian velocity at radius $r$ \citep{2015ApJ...804..101Y}.

The mass flux of the wind increases with radius, and would be comparable with inflow rate at the outer boundary of the hot accretion flow (i.e., truncation radius $r_{\rm tr}$) $\dot{M}_{\rm out}\simeq\dot{M}_{\rm in,net}(r_{\rm tr})$. 
Thus the wind could efficiently suppress the mass inflow, making the net accretion rate at the black hole event horizon much smaller than the inflow rate at $r_{\rm tr}$ ($\dot{M}_{\rm in,net}(r_{\rm EH})\ll\dot{M}_{\rm in,net}(r_{\rm tr})$).

\subsection{Sample Selection} \label{subsec:sample}

We retrieve the properties of all types of winds reported in low-luminosity AGN with X-ray observations from literature, summarized in Table. \ref{tab:prop}.
To obtain a reasonable sample size, we extend the Eddington ratio threshold to $R_{\rm Edd}\lesssim0.04$ for AGN with low-luminosity end.

For sources with multiple wind components reported, we select the component with the largest velocity to trace the wind as close to the central black hole as possible. This is because we are mainly interested in the wind launched from the accretion flow while the closest component would be less subject to the interaction with circumnuclear medium and represent the intrinsic wind properties.
We take the average of parameters for wind components with the largest velocity reported in multiple literature with the same observations, while keep as an independent record for that derived from different observations.
A total of 21 wind components discovered in 14 sources have been retrieved.
Properties of these winds are collected from \citet{2021NatAs...5..928S,2022ApJ...926..209S,2014MNRAS.441.2613L,2016MNRAS.457.3896L,2009ApJ...691..922M,2021A&A...652A.150M, 2022A&A...657A..77W,2005A&A...434..569S,2007MNRAS.379.1359M,2010ApJ...711..888A,2019A&A...625A..25M,2005ApJ...633..693K,2010A&A...521A..57T,2011ApJ...742...44T,2013MNRAS.430.1102T,2015ApJ...808..154R,2024MNRAS.532.3036M} and black hole mass are taken from \citet{2015PASP..127...67B}.

\begin{deluxetable*}{l|llllllllc}[htbp]
\tabletypesize{\scriptsize}
%\tabletypesize{\tiny}
\tablewidth{0pt}
\renewcommand{\arraystretch}{1.2}
\setlength{\tabcolsep}{4pt}
%\tablenum{1}
\tablecaption{Properties of wind in low-luminosity AGN sample\label{tab:prop}}
\tablehead{
\colhead{Source} & \colhead{Type}& \colhead{$M_{\rm BH}$} & \colhead{log($R_{\rm Edd}$)} & \colhead{$v_{\rm wind}$} & \colhead{log($R_{\rm max}$)} & \colhead{log($r_{\rm tr,SANE00}$)} & \colhead{log($r_{\rm tr,MAD98}$)} & \colhead{$\dot{M}_{\rm in}$} & \colhead{$\dot{M}_{\rm out}$ (min/max)}\\
\colhead{} & \colhead{} & \colhead{($10^7~\rm M_{\odot}$)}& \colhead{} & \colhead{($\rm km~s^{-1}$)} & \colhead{($\rm r_g$)} & \colhead{($\rm r_g$)} & \colhead{($\rm r_g$)} & \colhead{($\rm M_{\odot}~yr^{-1}$)} &  \colhead{($\rm M_{\odot}~yr^{-1}$)}\\
\colhead{(1)} & \colhead{(2)} & \colhead{(3)} & \colhead{(4)} & \colhead{(5)} & \colhead{(6)} & \colhead{(7)} & \colhead{(8)} & \colhead{(9)} & \colhead{(10)}
} 
%\colnumbers
\startdata 
M81* & HW & $7^{+2}_{-1}$ & $-4.64_{-0.11}^{+0.07}$ & $-2800^{+200}_{-200}$ & $3.7_{-0.2}^{+0.3}$ & $2.69_{-0.15}^{+0.15}$ & $3.66_{-0.15}^{+0.15}$ &  $0.00008$ & - / $0.002$\\
\hline
NGC\,7213 & HW & $8^{+16}_{-6}$ & $-3.0_{-0.6}^{+0.8}$ & $-1200_{-200}^{+100}$ & $3.9_{-0.4}^{+0.4}$ & $3.3_{-0.6}^{+0.6}$ & $4.3_{-0.6}^{+0.6}$ & $0.003$ & - / $0.08$\\
\hline
IRAS\,050278 & WA & ${7.2^{+0.7}_{-0.7}}^{*}$ & $-1.61_{-0.04}^{+0.05}$ & $-900_{-30}^{+600}$ & $10.1_{-0.2}^{+0.2}$ & $3.68_{-0.07}^{+0.07}$ & $4.65_{-0.07}^{+0.07}$ & $0.049$ & $0.017$ / $1952$ \\
\hline
\multirow{2}{*}{NGC\,3227} & WA & \multirow{2}{*}{$2.0^{+1.0}_{-0.4}$} & \multirow{2}{*}{$-1.73_{-0.18}^{+0.10}$} & $-1270_{-120}^{+20}$ & $3.85_{-0.05}^{+0.05}$ & $3.4_{-0.2}^{+0.2}$ & $4.3_{-0.2}^{+0.2}$ & \multirow{2}{*}{$0.01$} & $0.0016$ / $0.0037$ \\
{    } & WA & {    } & {  } & $-2060_{-170}^{+240}$ & $5.3_{-0.3}^{+0.3}$ & $2.9_{-0.2}^{+0.2}$ & $3.9_{-0.2}^{+0.2}$ & { } & $0.0126$ / $0.032$\\
\hline
\multirow{3}{*}{NGC\,5548} & WA & \multirow{3}{*}{$9.3^{+0.6}_{-0.6}$} & \multirow{3}{*}{$-2.6_{-1.1}^{+0.3}$} & $3610_{-270}^{+180}$ & $6.75_{-0.12}^{+0.12}$ & $2.47_{-0.07}^{+0.07}$ & $3.44_{-0.07}^{+0.07}$ & \multirow{3}{*}{$0.007$} & $0.0155$ / $9.35$\\
{    } & WA & {   } & {  } & $-1040_{-150}^{+150}$ & $6.32_{-0.17}^{+0.18}$ & $3.56_{-0.11}^{+0.11}$ & $4.53_{-0.11}^{+0.11}$ & {  } & $0.29$ / $0.96$ \\
{    } & WA & {   } & {  } & $-1180_{-150}^{+150}$ & $7.6_{-0.1}^{+0.1}$ & $3.45_{-0.10}^{+0.10}$ & $4.42_{-0.10}^{+0.10}$ & {  } & $0.05$ / $3.95$ \\
\hline
\multirow{2}{*}{3C59} & WA & \multirow{2}{*}{${79^{+8}_{-8}}^{*}$} & \multirow{2}{*}{$-1.41_{-0.04}^{+0.05}$} & $-3530_{-130}^{+130}$ & $7.85_{-0.06}^{+0.05}$ & $2.49_{-0.07}^{+0.07}$ & $3.46_{-0.07}^{+0.07}$ & \multirow{2}{*}{$0.85$} & $0.072$ / $716$ \\
{     } & WA & {   } & { } & $-1000^{+120}_{-120}$ & $6.32_{-0.06}^{+0.05}$ & $3.59_{-0.11}^{+0.10}$ & $4.56_{-0.11}^{+0.10}$ & {  } & $0.51$ / $12.1$ \\
\hline
3C382 & WA & ${115^{+12}_{-12}}^{*}$ & $-1.55_{-0.04}^{+0.05}$ & $-1530^{+370}_{-370}$ & $7.21_{-0.17}^{+0.12}$ & $3.22_{-0.16}^{+0.17}$ & $4.19_{-0.16}^{+0.17}$ &  $0.88$ & $0.041$ / $18$\\
\hline
\multirow{2}{*}{4C+74.26} & WA & \multirow{2}{*}{${417^{+42}_{-42}}^{*}$} & \multirow{2}{*}{$-1.70_{-0.04}^{+0.05}$} & $-1490_{-90}^{+90}$ & $7.10_{-0.06}^{+0.05}$ & $3.24_{-0.08}^{+0.08}$ & $4.21_{-0.08}^{+0.08}$ & \multirow{2}{*}{$2.3$} & $0.80$ / $254$\\
{     } & WA & {   } & {   } & $-3000_{-500}^{+500}$ & $6.06_{-0.07}^{+0.06}$ & $2.64_{-0.13}^{+0.13}$ & $3.60_{-0.13}^{+0.13}$ & {  } & $0.75$ / $88$\\
\hline
Mrk\,6 & WA & $12.7^{+1.1}_{-1.1}$ & $-1.84_{-0.04}^{+0.04}$ & $-4000_{-500}^{+500}$ & $6.69_{-0.07}^{+0.06}$ & $2.39_{-0.10}^{+0.11}$ & $3.36_{-0.10}^{+0.11}$ & $0.05$ & $0.041$ / $25.8$ \\
\hline
PKS\,2135-14 & WA & ${450^{+45}_{-45}}^{*}$ & $-1.74_{-0.04}^{+0.05}$ & $-1240_{-530}^{+530}$ & $7.4_{-0.4}^{+0.2}$ & $3.4_{-0.3}^{+0.3}$ & $4.4_{-0.3}^{+0.3}$ & $2.2$ & $0.14$ / $64$\\
\hline
\multirow{3}{*}{NGC\,4151} & WA & \multirow{3}{*}{$2.37_{-0.15}^{+0.17}$} & \multirow{3}{*}{$-1.64_{-0.03}^{+0.03}$} & $-490_{-50}^{+50}$ & $4.3_{-0.3}^{+0.2}$ & $4.21_{-0.08}^{+0.08}$ & $5.18_{-0.08}^{+0.08}$ & \multirow{3}{*}{$0.015$} & $1.16$ / $5.6$ \\
{    } & WA & {   } & {   } & $-370_{-40}^{+40}$ & $5.69_{-0.19}^{+0.19}$ & $4.46_{-0.09}^{+0.08}$ & $5.43_{-0.09}^{+0.08}$ & {  } & $1.26$ / $0.93$ \\
{    } & UFO & {   } & {   } & $-31800_{-2100}^{+2100}$ & \textless$3.25$ & $0.59_{-0.07}^{+0.07}$ & $1.55_{-0.07}^{+0.07}$ & {  } & $0.003$ / $0.04$\\
\hline
Mrk\,205 & UFO & $40_{-36}^{+358}$ & $-1.9_{-1.0}^{+1.0}$ & $-30000_{-1200}^{+1200}$ & \textless$2.42$ & $0.2_{-1.1}^{+1.1}$ & $1.1_{-1.0}^{+1.1}$ & $0.13$ & $0.63$ / $0.63$\\
\hline
Cygnus\,A & UFO & $250\pm70$ & $-1.90_{-0.11}^{+0.14}$ & $-19000^{+10000}_{-7000}$ & \textless$3.34$ & $1.0_{-0.3}^{+0.3}$ & $2.0_{-0.3}^{+0.3}$ & $0.9$ & $7.7$ / $247$ \\
\hline
NGC\,7582 & UFO & $1.3\pm0.3$ & $-1.9\pm0.1$ & $-85000_{-900}^{+900}$ & $\textless2.78$ & - & $0.69_{-0.15}^{+0.15}$ & $0.004$ & $0.023$ / $11.7$\\
\enddata
\tablecomments{
(1) Source name;
(2) Outflow type: {\it `HW'} for hot wind, {\it `WA'} for warm absorbers, {\it `UFO'} for ultra-fast outflow;
(3) Black hole mass. Errors are determined by 10\% uncertainties if no explicit error bars are provided in the literature for measurement marked by `*';
(4) Eddington Ratio;
(5) Wind velocity;
(6) Location of wind assuming its depth equals its distance to SMBH ($\Delta r=r$);
(7)-(8) Putative truncation radius assuming SANE or MAD type accretion flow surrounding SMBH with spin $a$ of 0 or 0.98.
(9) Mass inflow rate deduced from $L_{\rm bol}$;
(10) Lower / Upper limit of the mass outflow rate;
}
\end{deluxetable*}

We divide the collected wind components detected in the low-luminosity AGN into three categories.
Ultra-fast outflows (hereafter UFO) are characterized by blue-shifted absorption lines with large bulk velocity higher than $\gtrsim10^4\rm~km~s^{-1}$ \citep{2010A&A...521A..57T, 2023A&A...670A.182M}.
Hot winds (hereafter HW) driven by hot accretion flow with large opening angle have firstly been detected in M81* and NGC\,7213, traced by highly collisionally ionized iron emission lines with blue-/red-shift velocity of $\sim10^3\rm~km~s^{-1}$ in the hard X-ray band. Although the confirmed cases of hot wind remain scarce, gaining evidence indicates their prevalence in LLAGN.
Warm absorbers (hereafter WA) are mass outflows of ionized clouds characterized by blue-shifted absorption lines in soft X-ray of UV bands with typical velocity ranging $\sim10^2\rm~km~s^{-1}$ -- $\sim10^3\rm~km~s^{-1}$.

Assuming the observed wind originates from the outer edge ($r\simeq r_{\rm tr}$) of the hot accretion flow, the velocity of winds would remain almost constant as the wind freely propagates outward and before it encounters the interstellar medium \citep{2020ApJ...890...81C}.
We consider two extreme assumption SANE00 and MAD98.
Based on Eq. (\ref{eq:vSANE00}) and Eq. (\ref{eq:vMAD98}),
we deduce the putative truncation radius $r_{\rm tr}$ from the observed velocity of wind $v_{\rm wind}$, as recorded in Table \ref{tab:prop}. We neglect the projection effect caused by the viewing angle for simplicity.

We use $R_{\rm max}$ to describe the average location of wind determined through observation (either emission or absorption).
For HW detected with emission lines, $R_{\rm max}$ represents the radius where $50$\% of total luminosity of blue-shifted Fe XXVI line resides within (from Figure 2 in \citet{2024ApJ...970...48S}).
For WA and UFO, by combining the column density $N_{\rm H}=n_{\rm H}V_{\rm f}\Delta r$ and ionization parameter $\xi=\frac{L_{\rm ion}}{n_{\rm e}r^2}$ derived from spectral modeling in literature, while assuming the depth of wind is roughly equivalent to its distance to black hole $\Delta r\simeq r$, the wind location is determined by $R_{\rm max}=\frac{L_{\rm ion}V_{\rm f}}{\xi N_{\rm H}}$ \citep{2016MNRAS.457.3896L,2024MNRAS.532.3036M}.
Here we assume that volume filling factor is $V_{\rm f}\sim1$ and that the number density of hydrogen $n_{\rm H}$ are roughly equals to electron $n_{\rm e}$ in the wind for simplification.
The ionizing luminosity $L_{\rm ion}$ is over $13.6\rm~eV$--$13.6\rm~keV$ band.
The mass outflow rate at $R_{\rm max}$ thus can be calculated with $\dot{M}_{\rm max}=4\mu\pi m_{\rm p}L_{\rm ion}\xi^{-1}v_{\rm wind}C_{\rm f}$, where the average relative atomic mass $\mu\simeq1.4$ and a covering factor $C_{\rm f}\sim0.5$ is assumed \citep{2024MNRAS.532.3036M}. We perform bootstrapping to determine the range of 90\% confidence level of the derived $R_{\rm max}$.
We have also estimated the lower limit of the mass outflow rate $\dot{M}_{\rm min}=8\mu\pi m_{\rm p}GM_{\rm BH}N_{\rm H}v_{\rm wind}^{-1}C_{\rm f}$ for UFO and WA, assuming the wind is launched where its velocity exceed the local escape velocity.
We have to emphasize that this estimation suffers from large uncertainties.

The bolometric luminosity of AGN implies the energy released by the accretion material at the event horizon of the central black hole. The mass inflow rate at the event horizon would be estimated by $\dot{M}_{\rm in}=\frac{L_{\rm bol}}{\epsilon c^2}$, where $\epsilon$ refers to the radiation efficiency. For radiatively inefficient hot accretion flows, $\epsilon$ decreases with the decreasing of the Eddington ratio $R_{\rm Edd}\equiv\frac{L_{\rm bol}}{L_{\rm Edd}}$ and falls below that of the standard thin disk ($\epsilon_{\rm SSD}\simeq 0.1$) in the low-luminosity regime. In this paper, we adopt the radiation efficiency $\epsilon$ from the theoretical $\epsilon$ -- $\frac{\dot{M}_{\rm in}}{\dot{M}_{\rm Edd}}$ correlation provided in \citet{2012MNRAS.427.1580X}, assuming the fraction from viscous heating is $\delta\sim0.5$. Their results are consistent with \citet{2017MNRAS.468.1398S}.
The Eddington accretion rate is defined as $\dot{M}_{\rm Edd}\equiv10L_{\rm Edd}/c^2$. So the ratio between the net inflow rate at event horizon and the Eddington accretion rate are linked to the Eddington ratio $R_{\rm Edd}\equiv\frac{\epsilon\dot{M}_{\rm in}}{0.1\dot{M}_{\rm Edd}}$.
Thus given any Eddington ratio $R_{\rm edd}$ we would be able to precisely determine the net inflow rate at event horizon.

\section{Results} \label{sec:stat}

\begin{figure*}[htbp]
\centering
\includegraphics[width=0.48\linewidth]{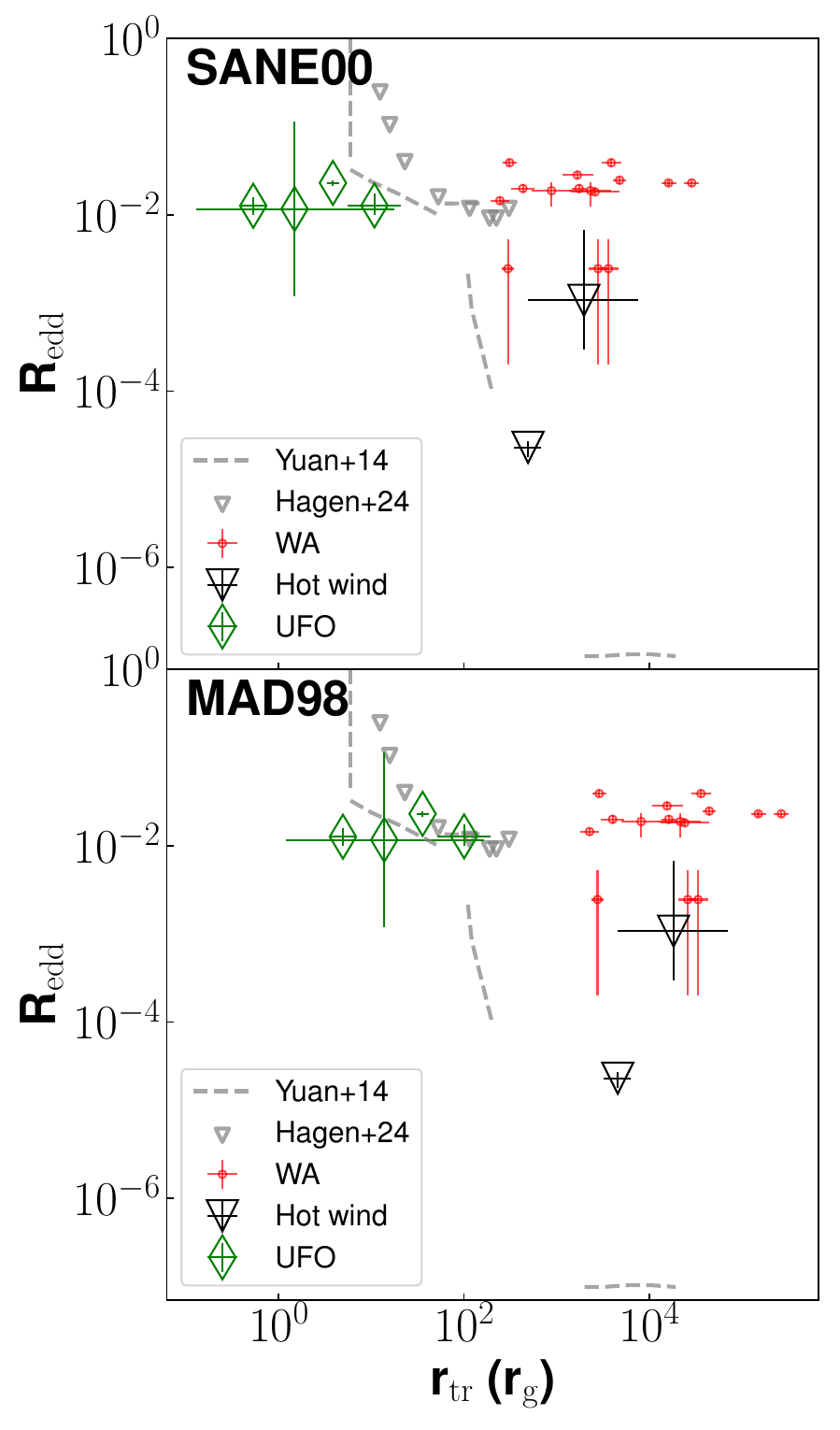}
\includegraphics[width=0.475\linewidth]{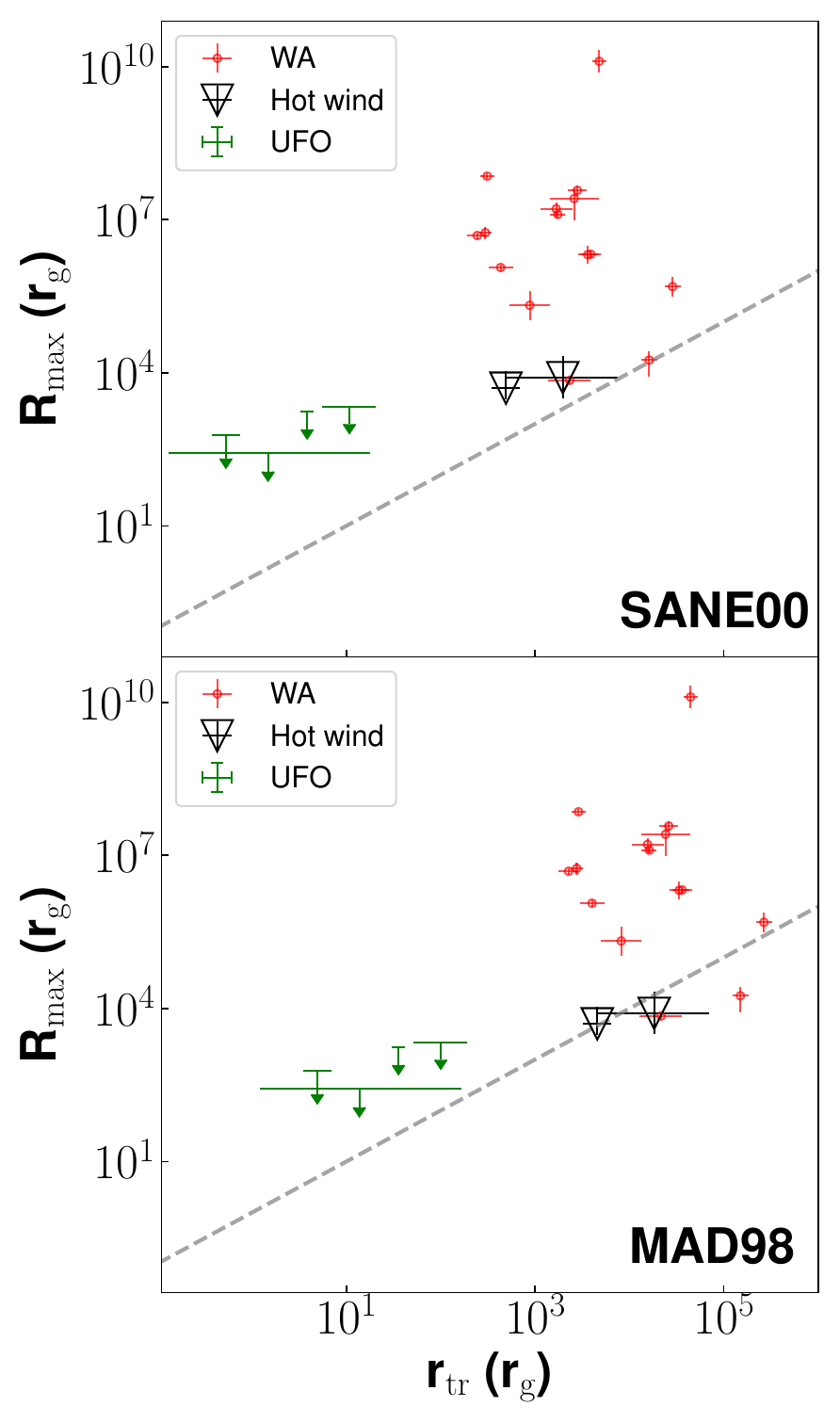}
\caption{
{\it Left panel:} Putative truncation radius $r_{\rm tr}$ of the selected sample as a function of Eddington ratio $R_{\rm Edd}$, assuming driven by SANE00 (top panel) and MAD98 (bottom panel).
Grey dashed lines and triangles present the fiducial value from SED fitting given by previous studies for comparison, taken from \citet{2014ARA&A..52..529Y} and \citet{2024MNRAS.534.2803H} respectively.
Green arrows denote the Ultra-fast outflows (UFO). Black triangles describe wind detected with highly ionized emission lines (Hot wind). Red diamonds represent the warm absorbers (WA).
Putative $r_{\rm tr}$ of UFO and Hot wind in MAD98 case is more consistent with the fiducial relation.
{\it Right panel:} Wind location $R_{\rm max}$ determined from spectral modeling compared with putative $r_{\rm tr}$ for SANE00 (top) and MAD98 (bottom).
The grey dashed line marks where $r_{\rm tr}=R_{\rm max}$.
\label{fig:Rtr_comb}
}
\end{figure*}

We compare the putative truncation radii estimated from the wind velocity using the theory of wind in our sample as a function of Eddington ratio with the fiducial correlation taken from \citet{2014ARA&A..52..529Y}, the result is shown in the left panel of Fig. \ref{fig:Rtr_comb}.
The fiducial truncation radii were determined by broadband SED fitting with RIAF models \citep{2004ApJ...612..724Y}.
LLAGNs with UFO are roughly more consistent with the MAD98 case.
M81* and NGC\,7213 where hot winds have been confirmed through highly ionized iron emission lines, together with Mrk\,205 and NGC\,4151 with UFOs generally follow the trend that $r_{\rm tr}$ increases as the accretion power decreases. This trend between $R_{\rm edd}$ and $r_{\rm tr}$ deduced from black hole wind is quantitatively consistent with the theoretical prediction from spectral modeling.
While $r_{\rm tr}$ estimated from WA components are in general larger than the fiducial values.

The right panel of Fig. \ref{fig:Rtr_comb} shows the comparison between $r_{\rm tr}$ and wind location $R_{\rm max}$ deduced from photoionization modeling. We can see from the figure that, the estimated $r_{\rm tr}$ for both UFO and hot wind are in line with $R_{\rm max}$, especially in the case of MAD with high BH spin.
In contrast, WAs lie above the line where $R_{\rm max}=r_{\rm tr}$.
For UFO and hot wind, the detection location is close to the black hole and more specifically, close to the truncation radius.
So the detected wind velocity would be roughly consistent with the initial velocity of wind at its launching point. Our assumption is valid that the wind velocity originated from truncation radius remains almost constant before encountering ISM \citep{2020ApJ...890...81C}.
But for WA, the detection location seems far away from the wind launching region. 
One possible scenario is that although in LLAGNs outflows are driven mainly from $r_{\rm tr}$, the detected WAs trace the wind components located at a larger radius ($\gtrsim 10^5\rm~r_g$) away from their origin points. 
Only at such large distances, the ionization state of WA is lower and ``suitable'' to be detected in the soft X-rays. As the distance is further away, WAs may be opt to experience deceleration through interacting with interstellar medium (ISM), resulting in a larger putative truncation radius according to Eq. (\ref{eq:vSANE00}) - (\ref{eq:vMAD98}) with smaller wind velocity than expected.
Since WAs may not trace the launching site of winds in the LLAGN, the intrinsic truncation radius of these AGN would be much smaller than deduced.
And as the distance increases, the initially driven wind is expected to entrain more and more mass\citep{2017ApJ...837..149G}.
So the mass outflow rate estimation of these WAs would not reflect the intrinsic black hole wind.

\begin{figure}[hbtp]
\centering
\includegraphics[width=0.98\linewidth]{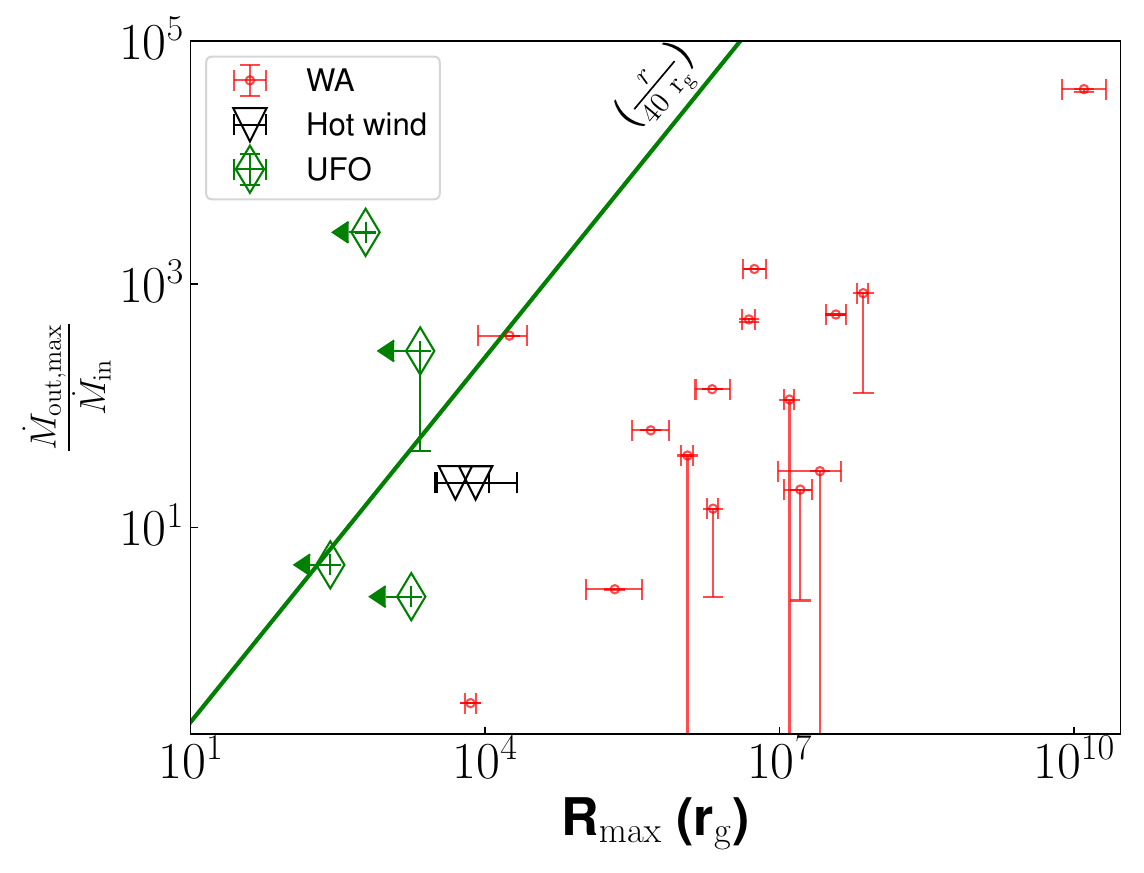}
\caption{Ratio between the observationally derived mass flux of wind $\dot{M}_{\rm out,max}$ at $R_{\rm max}$ and the black hole accretion rate $\dot{M}_{\rm in}=\frac{L_{\rm bol}}{\epsilon c^2}$ as a function of wind location $R_{\rm max}$.
The vertical solid line below each data point indicates the estimated lower limit of the mass flux $\dot{M}_{\rm out,max}$ for each detected wind component. 
The green solid line denotes the theoretical prediction for mass flux of wind, $\frac{\dot{M}_{\rm out,max}}{\dot{M}_{\rm in}}=\frac{r}{40~\rm r_g}$ \citep{2015ApJ...804..101Y}.
Other captions are similar to Fig. \ref{fig:Rtr_comb}
\label{fig:Mout_v1}
}
\end{figure}

Wind driven by hot accretion flow is expected to carry a substantial amount of material away. The radial profile of its mass outflow rate can be described by $\dot{M}_{\rm out,max}=\dot{M}_{\rm in}(\frac{r}{40\rm~r_g})$ according to numerical simulations \citep{2015ApJ...804..101Y}.
$\dot{M}_{\rm in}$ refers to the mass accretion rate at event horizon (See in Sec. \ref{subsec:sample}).
We investigate the ratio between the wind mass flux and black hole accretion rate (also known as the mass loading factor) for our sample. The results are shown in Figure \ref{fig:Mout_v1}.
For UFO together with hot wind, the ratio is consistent with the theoretical estimation, while all the WAs are systematically offset from the expected correlation.
\citet{2024A&A...686A..36F} collects a sample of wind components mostly detected in AGN with intermediate and high accretion rate and suggests a negative power-law correlation between the mass loading factor and the Eddington ratio. 
However in our LLAGN wind sample, we do not find such a strong correlation and the mass loading factor are on average larger than their high accretion rate counterparts.
This adds further support to a different wind launching mechanism between hot mode and cold mode accretion.

\section{Detailed Analysis on the prototype M81*}\label{sec:M81case}

Among all the selected sample, M81* is the closest archetypal LLAGN and has the most abundant multi-wavelength observations over the past two decades.
It is also one of the only two LLAGNs where evidence of hot wind has been confirmed by high-resolution X-ray spectroscopy. In this section, we take M81 as an example, to compare the observational results and theoretical prediction of hot wind.

\subsection{Truncation radius}
There are several ways to determine the truncation radius.
By fitting the broadband spectral energy distribution of M81* with ADAF + thin disk model, \citet{1999ApJ...525L..89Q} found the truncation radius would be around $10^2\rm~r_g$ and \citet{2014MNRAS.438.2804N} gave an updated estimation of $360\rm~r_g$.
The second way relates to the fluorescent $6.4\rm~keV$ iron K$\alpha$ line. Assuming it is generated from the Keplerian rotating cold outer thin disk illuminated by primary emission from hot X-ray corona,
from its line width $\sigma_{\rm v}$ 
we would determine $r_{\rm tr}\gtrsim4.6\times10^2-1.9\times10^4\rm~r_g$ for M81* \citep{2007ApJ...669..830Y,2021NatAs...5..928S}. 
Real $r_{\rm tr}$ could be even larger than the above estimation since
\citet{2018MNRAS.476.5698Y} concludes that fluorescent Fe K$\alpha$ line in M81* is produced from optically thin material within the $r_{\rm tr}$.
The third way is from the reflection fraction $R_{\rm refl}$.
The weak reflection component in the X-ray spectra of M81* indicates an lower limit of the truncation radius $r_{\rm tr}\gtrsim(3-9)\times10^2\rm~r_g$ \citep{2018MNRAS.476.5698Y,2021NatAs...5..928S}.
The derived truncation radius of these various method are roughly consistent with that predicted by hot accretion flow theory \citep{2004ApJ...612..724Y}.

\subsection{Velocity of Wind}

As we have introduced in \S2.1, the theory of wind  \citep{2012ApJ...761..130Y,2015ApJ...804..101Y} predicts that the wind speed should be proportional to the Keplerian speed at the wind launching radius (in this case, the truncation radius).
Using the above-mentioned observationally estimated truncation radius, this theory predicts that the velocity of wind would be $(0.46-3.1)\times10^3\rm~km~s^{-1}$ for SANE00 and $(1.4-9.6)\times10^3\rm~km~s^{-1}$ for MAD98.
The observed wind velocity of M81* derived from X-ray spectroscopy is $\sim2.8\times10^3\rm~km~s^{-1}$. This value is well within the predicted range.

Specifically, in the customized hot wind simulation for M81* $r_{\rm tr}=3000~\rm r_{\rm g}$ is adopted in \citet{2021NatAs...5..928S}. The corresponding Keplerian velocity of this radius is $v_{\rm k}\sim5.4\times10^3\rm~km~s^{-1}$. Assuming in M81 the accretion flow is MAD and the black hole is rapidly spinning, the wind theory predicts a speed of $v_{\rm wind}\sim3.5\times10^3\rm~km~s^{-1}$.
The observed wind velocity is well consistent with this prediction. 

\subsection{Mass accretion rate at the event horizon}
The bolometric luminosity of M81* is $\sim 2.3\times 10^{-5}L_{\rm Edd}$. 
Using the result of radiative efficiency as a function of the mass accretion rate at the black hole horizon presented in \citep{2012MNRAS.427.1580X} (as described in Sec. \ref{subsec:sample}), we can obtain the net accretion rate at the event horizon of black hole $\dot{M}_{\rm in}(r_{\rm EH})=\frac{L_{\rm bol}}{\eta c^2}\sim8.5\times10^{-5}\rm~M_{\rm\odot}~yr^{-1}$.

\subsection{Mass flux of wind}
The theory of hot wind has predicted that most of the accretion material should be lost in wind and only a small fraction can finally reaches the black hole. In other words, the mass flux of wind should be roughly equal to the mass flux at the outer boundary of the hot accretion flow, i.e., the truncation radius \citep{2012ApJ...761..129Y,2012ApJ...761..130Y,2015ApJ...804..101Y}. In this subsection, we use observational data of M81 to examine this prediction. 

\citet{2011MNRAS.413..149S} estimated a mass inflow rate of $\dot{M}_{\rm in}(17\rm~pc)\sim4\times10^{-3}\rm~M_{\odot}~yr^{-1}$ from the observed kinematics of ionized gas within $\sim17\rm~pc~(5\times10^6\rm~r_g)$.
Since this location is much larger than the estimated truncation radius above, these ionized gas should supply the outer accretion thin disk. Wind exists in the thin disk and the  mass flux of wind is  about half of the inflow rate in the disk (see Fig. 7 in \citet{2022MNRAS.513.5818W}). So the mass inflow rate left at the innermost radius of the thin disk, i.e., the truncation radius, would be 
$\dot{M}_{\rm in,net}(r_{\rm tr})\sim0.5\dot{M}_{\rm in}(17\rm~pc)\sim2\times10^{-3}\rm~M_{\odot}~yr^{-1}$.

By comparing the observed highly ionized iron emission lines with the hot wind modeling from customized numerical simulations, \citet{2021NatAs...5..928S} reported a mass outflow rate $\dot{M}_{\rm out}\sim2\times10^{-3}\rm~M_{\odot}~yr^{-1}$ of hot wind in M81*. This is in excellent agreement with the prediction of theory of wind.

\subsection{Radial profile of hot accretion inflow}

Combining the mass accretion rates at the black hole event horizon and at the truncation radius obtained in the previous two subsections, we can estimate the index $s$ of the radial profile of accretion rate $\dot{M}_{\rm in,net}(r_{\rm s})=\dot{M}_{\rm in}(r_{\rm tr})(\frac{r_{\rm EH}}{r_{\rm tr}})^s$ \citep[e.g.,][]{2012ApJ...761..129Y}. It is found to be $s\sim0.43$ using $r_{\rm tr}=3000\rm~r_{\rm g}\equiv1500~r_{\rm s}$ in M81*.
This is well consistent with the theoretical prediction \citep{2012ApJ...761..129Y,2021ApJ...914..131Y}. Specifically,
\citet{2021ApJ...914..131Y} predict the index $s$ to be $0.42$ for the case of MAD and black hole spin $a=0.98$, which matches with the obtained $s$ in M81* quite well.

The above result suggests that the accretion flow in M81 is MAD and the black hole should be rapidly spinning. This is consistent with the fact that a jet exists in M81, assuming the jet formation theory proposed by \citet{1977MNRAS.179..433B}. 
We note that this theory of jet formation recently has obtained new strong observational evidence. \citet{2024SciA...10N3544Y} have performed GRMHD numerical simulation of black hole accretion and jet formation. Assuming that magnetic re-connection in the jet can efficiently accelerate electrons, with detailed calculation of electron acceleration and radiative transfer in the jet, they obtained the jet morphology, including the elongated structure, the limb-brightening feature and the jet width as a function of distance to the black hole. Comparing these features with observations of the jet in M87, they find that the BZ jet can nicely reproduce all these observed features while other jet models, such as BP model, can not. Moreover, the mode of the accretion low should be MAD rather than SANE, and the black hole spin should be large. 
Our current results strongly suggest that, like M87, the accretion flow in M81* should also be MAD and the black hole spin in M81 should also be large.

\subsection{Jet-to-wind power ratio}

In the case of M81*, the kinetic power of hot wind is estimated to be $\sim2\times10^{40}\rm~erg~s^{-1}$ \citep{2021NatAs...5..928S}. 
Radio emission from M81* is highly variable,
with an average $1.4$ GHz flux density of $0.09$-$0.62\rm~Jy$ \citep{1992ApJS...79..331W,1998AJ....115.1693C}.
Based on the empirical relation of jet mechanical power $L_{\rm mech}=7\times10^{36}f(L_{\rm 1.4~GHz}/10^{25}\rm~W~Hz^{-1})^{0.68}\rm~W$ and assuming the calibration factor $f=4$ \citep{2014ARA&A..52..589H}, we roughly estimate the mechanical power of the jet in M81* would be around $(1.39-5.22)\times10^{41}\rm~erg~s^{-1}$.
The kinetic power carried by jet is of around several times that carried by the hot wind in M81*.
This is consistent with the prediction for an MAD around an extremely spinning black hole, where the ratio of kinetic energy between jet and hot wind would be $\sim4$ \citep{2021ApJ...914..131Y}.

In summary, we find good agreement between theory of wind launched from hot accretion flow and observations, including the speed and mass flux of the wind, the power-law index of the radial profile of inflow rate, and the jet-to-wind power ratio.

\section{Summary and Discussion}\label{sec:sum}

In this paper, we collect a sample of outflows (including hot wind, UFOs and WAs) detected in low-luminosity AGN ($R_{\rm Edd}\lesssim0.04$).
The inner accretion disk of these SMBHs with low accretion rate could be truncated and replaced by radiatively inefficient hot accretion flow.
We compare the theoretical predictions of wind driven by such hot accretion flow with observed wind properties.
Furthermore we analyze the multiple kinetic properties of inflow and outflow detected in M81* reported by multi-wavelength observations.
We take this closest prototypical LLAGN as a test case for the state-of-art theory of wind launched from hot accretion flows. Our main results are as follows.

\begin{itemize}

\item We have compared the truncation radii obtained from wind velocity based on  the theory of wind in our sample with the values of truncation radius obtained from other constraints. Good consistency is found, providing strong support to the theory of wind.  

\item We investigate the ratio between the wind mass flux and black hole accretion rate for our sample. 
For UFO together with hot wind, the ratio  is consistent with the theoretical prediction.

\item{For M81, using the truncation radius $r_{\rm tr}=3000 r_g$ obtained from broadband fitting, reflection fraction and Fe K$\alpha$ lines, the theory of wind predicts a speed of $\sim 3.5\times10^3\rm~km~s^{-1}$, very similar to the observed value of $\sim2.8\times10^3\rm~km~s^{-1}$.

\item 
The spectroscopy deduced outflow rate of the hot wind in M81* is roughly equal to the net inflow rate of warm ionized gas, which should be the inflow rate at the truncation radius. This is consistent with the theoretical  expectation. }

\item{The observed inflow rates at event horizon $r_{\rm EH}$ and at truncation radius $r_{\rm tr}$ suggest that the value of power index in the radial profile $\dot{M}_{\rm in,net}(r_{\rm s})=\dot{M}_{\rm in}(r_{\rm tr})(\frac{r_{\rm EH}}{r_{\rm tr}})^s$ $s\sim0.43$. This is consistent with the predicted value of $s\sim0.42$ for the case of MAD type hot accretion flow around an extremely spinning black hole.
}
\end{itemize}

%Overall, observational evidence of outflows detected in our LLAGN sample is consistent with theoretical predictions of accretion physics for hot accretion flow. We find accretion flow around M81* favors MAD and the the central black hole would be extremely spinning. This is further supported by the jet power in M81* is several times larger than kinetic power of hot wind as expected in extremely spinning black hole.
%So do the other UFOs and confirmed hot wind case in our selected sample. From the $r_{\rm tr}$ derived from wind velocity, both confirmed hot winds and reported UFOs are more consistent with driven by MAD type accretion flow around fast-spinning BH in the low-luminosity end. M87, a LLAGN with strong jet, has also been reported to be a Kerr-black hole with MAD type accretion flow \citep{2019ApJ...875L...6E,2022ApJ...924..124Y}.
%We suggest there may be potential association between detectable outflows (jets / winds) and high-spin black hole with MAD type accretion flow in LLAGN, which still requires confirmation from future observations.

We must emphasize that the wind velocity derived from observations may be subject to the projection effect of the viewing angle and wind properties like $\xi$, $N_{\rm H}$, and $R_{\rm max}$ that derived from photoionization modeling have large uncertainties. 
More attention should be paid to the outflows in the low-luminosity end of AGNs to achieve a larger and more accurate sample with the help of the new generation of X-ray telescopes like XRISM.

%% The "ht!" tells LaTeX to put the figure "here" first, at the "top" next
%% and to override the normal way of calculating a float position

%% Also note that the akcnowlodgment environment does not support long amounts of text. If you have a lot of people and institutions to acknowledge, do not use this command. Instead, create a new \section{Acknowledgments}.
\begin{acknowledgments}
We would like to thank Suoqing Ji and Defu Bu for helpful discussions.
F.S. is supported in part by the China Postdoctoral Science Foundation (grants 2022TQ0354 and 2022M723279).
F.Y. is supported by Natural Science Foundation of China (grants 12133008, 12192220, 12192223, and 12361161601) and the China Manned Space Project (CMS-CSST-2021-B02). 
\end{acknowledgments}


\vspace{5mm}
%\facilities{HST(STIS), Swift(XRT and UVOT), AAVSO, CTIO:1.3m,


%\software{astropy \citep{2013A&A...558A..33A,2018AJ....156..123A}}



%\appendix

%\section{Appendix information}


\bibliography{sample631}{}
\bibliographystyle{aasjournal}


\end{CJK*}
\end{document}